\documentclass[final,5p,times,number]{elsarticle}
\biboptions{sort&compress}

\usepackage{amssymb}
\usepackage{amsmath}
\usepackage{lipsum}
\usepackage{float}
\usepackage{ulem}
\usepackage[utf8]{inputenc}
\usepackage[colorlinks,linkcolor=blue,anchorcolor=blue,citecolor=blue,urlcolor=blue]{hyperref}

\journal{Physica A (accepted)}

\begin{document}

\begin{frontmatter}

\title{Phase behaviors and dynamics of active particle systems in double-well potential}

\author{Lu Chen$^{a,b}$ }
\affiliation{organization={College of Physics, Changchun Normal University},
            city={Changchun},
            postcode={130032},
            state={Jilin},
            country={China}}

\author{Baopi Liu$^{b}$ }
\affiliation{organization={Complex Systems Division, Beijing Computational Science Research Center},
            city={Beijing},
            postcode={100193},
            state={},
            country={China}}

\author{Ning Liu$^{c*}$}

\affiliation{organization={School of Mathematics and Physics, Anqing Normal University},
            city={Anqing},
            postcode={246133},
            state={Anhui},
            country={China}}

\begin{abstract}

In this study, we investigate the behaviors and dynamics of self-propelled particles with active reorientation (AR) in a double-well potential. We explore the competition between AR and external potentials, revealing that self-propelled particles exhibit flocking and clustering behaviors in an asymmetric potential trap. Through molecular dynamics simulations, we obtain a phase diagram that illustrates flocking behavior as a function of active reorientation and potential asymmetry. We compare the responses of inactive and active particles to the potential, finding that active reorientation significantly increases aggregation on one side of the asymmetric potential well. Additionally, by calculating the mean squared displacement and scaling exponent, we identify distinct diffusion regimes. Our findings demonstrate that active particles with active reorientation are more sensitive to variations in double-well potentials.
\end{abstract}


\end{frontmatter}

\section{Introduction}
\label{introduction}
The self-propulsion of active particles is very common in microorganisms. Systems composed of such particles with sustained self-propulsion capabilities exhibit a wealth of phenomena, such as phase separation~\cite{cates2015motility,de2021diversity,ginot2018aggregation}, flocking~\cite{cavagna2015flocking,bertrand2022diversity,chardac2021topolog,morin2015collective} and synchronization~\cite{casiulis2022emergent,levis2019activity}. In all these systems, interactions occur only between particles without any external forces. However, if active particles are exposed to an external field, it will inevitably affect their trajectories and collective dynamics.

There has been some research on the motion of active particles in an external field~\cite{morin2018flowing,caprini2022active,takatori2014swim,benjamin2022current}. Potentials can be used to create reconfigurable materials, such as guiding colloidal particle assembly~\cite{wu2020active,maloney2020clustering}. Furthermore, recent studies on the collective dynamics of active Brownian particles have revealed the impact of gravity~\cite{bickmann2022active}, external magnetic fields~\cite{kaiser2020directing,zakharchenko2018magnetic} and harmonic potential fields ~\cite{caprini2022active,abdoli2021stochastic,ebeling2005stochastic}. Among various types of potentials, the double-well potential plays a significant role in several physical systems~\cite{zhou2021stochasticity,marchenko2014particle,militaru2021escape}. In the work of Marchenko et al., they reveal that the underdamped motion of particles in space-periodic potentials is equivalent to overdamped motion within velocity space, resembling an effective double-well potential~\cite{marchenko2014particle}. In the work of Militaru et al., they study the effect of non-conservative forces on the escape dynamics of active particles in non-equilibrium systems by using a bistable optical potential trap to capture silica nanoparticles, finding an optimal correlation time that maximizes the escape rate~\cite{militaru2021escape}. Caprini et al. investigate the escape behavior of active particles in a double-well potential, revealing an optimal persistence time that minimizes the escape time and showing that repulsive interactions between active particles lead to correlated escape, a mechanism absent in systems without active particles~\cite{caprini2021correlated}. The double-well potential models these transitions, representing distinct functional states based on external signal strength or duration~\cite{li2017barrier}.

Recent work~\cite{das2024flocking} by Das et al. explored a model based on long-range dipolar interactions and non-reciprocal torques, where anisotropic interactions between hemispherical particles induce flocking behavior. In contrast, in our previous work~\cite{chen2023molecular}, we proposed a new active model consisting of self-propelled particles with active reorientation (AR). In addition to translational motion, the particles can also adjust their orientation through active reorientation. This concept of active reorientation is inspired by animal groups in nature~\cite{reynolds1987flocks,kasmuri2020human}, which actively adjust their orientation to avoid collisions when encountering obstacles. Our study is built on an AR theoretical framework inspired by experimental observations, investigating contact-induced reorientation driven by repulsive forces when particles overlap. Although both models exhibit flocking phenomena, they differ significantly in terms of particle interaction mechanisms, rotational dynamics, non-reciprocity and symmetry breaking, as well as experimental motivation and the range of applicable systems.

In this paper, we focus on the phase behaviors and dynamics of self-propelled particles with AR in a double-well potential, exploring the competition between AR and the external potential. First, we introduce the active system and the choice of potential. We then explore the relationship between phase behaviors (such as flocking and clustering) in this system and the parameter of asymmetry. Next, we study the dynamics of the system by examining the time dependence of particle number flow in symmetric and asymmetric wells, and by comparing the effects of the double-well potential in systems with and without active reorientation. Finally, we analyze the mean squared displacements (MSDs) to identify different diffusion regimes. The last section presents our conclusions.

\section{Theoretical Model}
Here, we consider an active particle model based on previous work~\cite{chen2023molecular,chen2023initial}. The system consists of \( N \) self-propelled disks in a two-dimensional square box with side lengths \( L_x \) (along the \( x \)-axis) and \( L_y \) (along the \( y \)-axis), both equal to \( L \). Each disk has a diameter \( d \) and is modeled as an active Brownian particle. The position of disk \( i \) is denoted by \( \boldsymbol{r}_i \), and its propulsion direction is represented by the orientation \( \theta_i \) of the polar vector \( \boldsymbol{p}_i = (\cos \theta_i, \sin \theta_i) \). The particle dynamics are governed by the following overdamped Langevin equations:

\begin{eqnarray}
\dot{\boldsymbol{r}}_i = v_0 \boldsymbol{p}_i \prod_{j=1, j \neq i}^N \varepsilon(r_{ij} - d) + \mu \boldsymbol{F}_i^{\text{in}} + \gamma \boldsymbol{F}_i^{\text{ex}} + \zeta_i(t), \label{eq-1}
\end{eqnarray}

\begin{eqnarray}
\dot{\theta}_i = \sum_{j=1, j \neq i}^N \alpha \delta(t) \varepsilon(d - r_{ij}) \frac{\boldsymbol{r}_{ij} \times \boldsymbol{p}_i}{|\boldsymbol{r}_{ij} \times \boldsymbol{p}_i|} \cdot \hat{\boldsymbol{z}} + \xi_i(t). \label{eq-2}
\end{eqnarray}

The first equation describes the translational motion of particle \( i \). The propulsion velocity has magnitude \( v_0 \), and \( \boldsymbol{r}_{ij} = \boldsymbol{r}_i - \boldsymbol{r}_j \) is the relative position vector between particles \( i \) and \( j \), with \( r_{ij} = |\boldsymbol{r}_{ij}| \). The unit vector pointing from particle \( j \) to \( i \) is \( \hat{\boldsymbol{r}}_{ij} = \boldsymbol{r}_{ij}/r_{ij} \). The Heaviside step function \( \varepsilon(x) \) equals 1 if \( x > 0 \), and 0 otherwise. The term \( \boldsymbol{F}_i^{\text{in}} = \sum_{j=1, j \neq i}^N \boldsymbol{F}_{ij} \) represents the repulsive forces, where \( \boldsymbol{F}_{ij} = k(d - r_{ij})\hat{\boldsymbol{r}}_{ij} \) for \( r_{ij} < d \), and 0 otherwise. Here, \( k \) is the interaction strength. The external force \( \boldsymbol{F}_i^{\text{ex}} = -\nabla U \) arises from an asymmetric potential \( U(x, y) \), and \( \gamma \) is the damping coefficient. The second equation governs the rotational motion of particle \( i \). \(\sum_{j=1, j \neq i}^N \alpha \delta(t) \varepsilon(d - r_{ij}) \frac{\boldsymbol{r}_{ij} \times \boldsymbol{p}_i}{|\boldsymbol{r}_{ij} \times \boldsymbol{p}_i|} \cdot \hat{\boldsymbol{z}}\sum_{j=1, j \neq i}^N \alpha \delta(t) \varepsilon(d - r_{ij}) \frac{\boldsymbol{r}_{ij} \times \boldsymbol{p}_i}{|\boldsymbol{r}_{ij} \times \boldsymbol{p}_i|} \cdot \hat{\boldsymbol{z}}\) describe the function of AR, where \( \alpha\) represents the strength of AR, and \( \hat{\boldsymbol{z}} \) is a unit vector normal to the plane. AR only occurs when there is an overlap between particles. Notably, \(\alpha\) is identical for each particle. The Gaussian white noises \( \zeta_i(t) \) and \( \xi_i(t) \) satisfy \( \langle \zeta_i(t) \rangle = 0 \), \( \langle \xi_i(t) \rangle = 0 \), \( \langle \zeta_i(t) \zeta_j(t') \rangle = 2D_T \delta_{ij}\delta(t - t') \), and \( \langle \xi_i(t) \xi_j(t') \rangle = 2D_R \delta_{ij}\delta(t - t') \), with \( D_T \) and \( D_R \) denoting the translational and rotational diffusion coefficients, respectively. In the numericals we have scaled lengths with the diameter $d$ of the particles and times with \(1/(\mu k)\).

\begin{figure}[t]%
\centering
\includegraphics[width=0.5\textwidth]{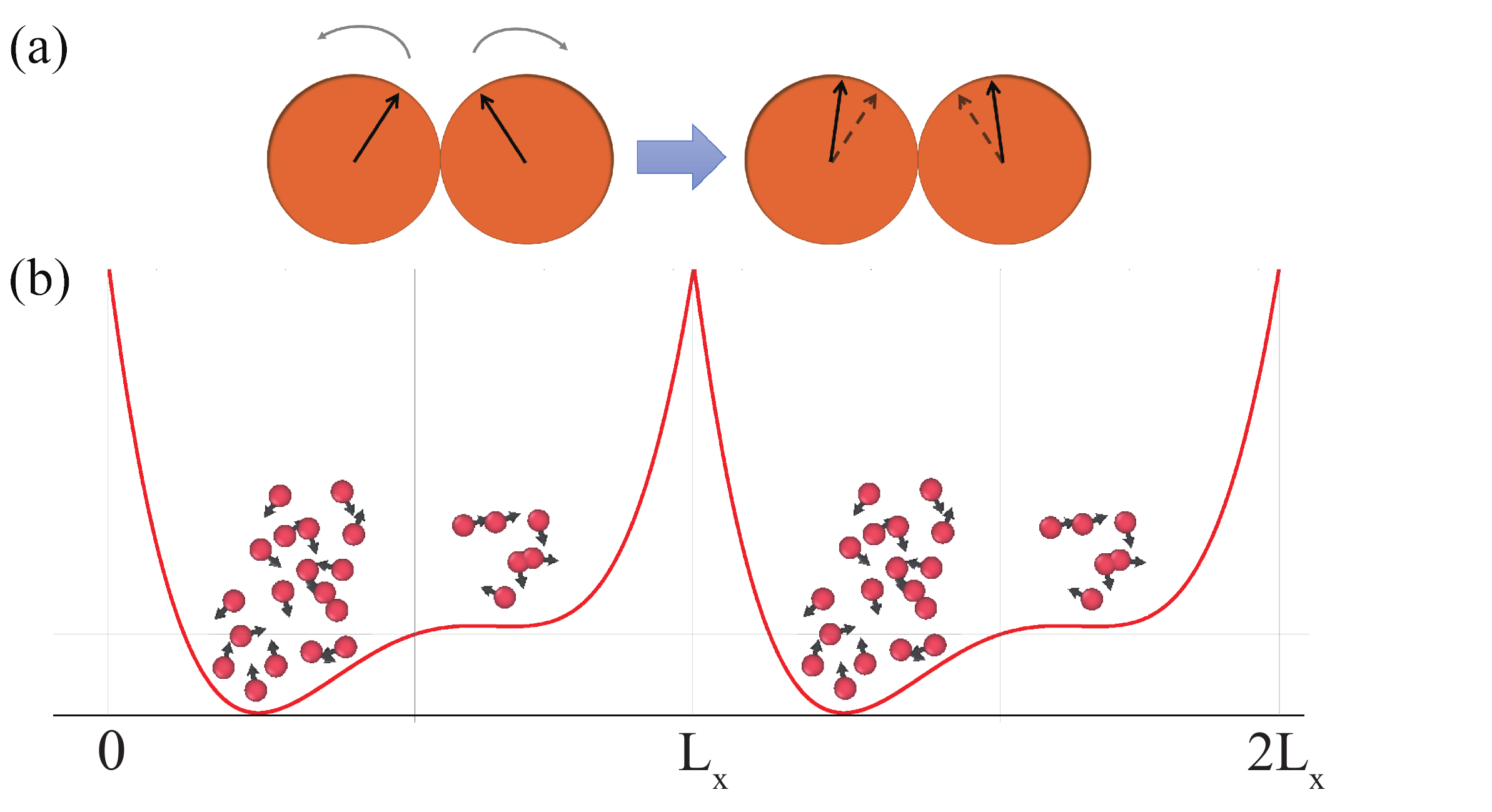}
\caption{Active particles in double-well potential.
The orange circles schematically show the state of particles before and after overlapping due to active reorientation. The red curve below presents the projection of the asymmetric potential $U(x,y)$ in the x-direction.}
\label{potential-illustrate}
\end{figure}

The external potential \( U(x, y) \) is quasi-one-dimensional,  expressed as :

\begin{equation}
U(x, y) = a(z^4 - z^2) + c z + b \bar{z}^2,
\end{equation}
where \( z = k(x + y) \), \( \bar{z} = k(x - y) \), $k$ is a dimensionless scaling factor. Here we set $k$ as 1. \( a \) and \( b \) control the strength of the potential, and \( c \) introduces asymmetry. U(x,y) is normalized in units of $k_B \mathcal{T}$. This potential is illustrated in Fig.\,\ref{potential-illustrate}. In this article, we primarily focus on the impact of potential asymmetry on the system, as the following analysis only varies the \( c \) parameter.

Simulation parameters include \( d = 1 \), \( a = 1/30 \), \( b = 1/1.5 \), periodic boundary conditions, a time step of \( 10^{-2} \), and a total integration time exceeding \( 10^6 \). Tests confirmed that after \( 10^6 \) steps, particle distribution and collective behavior no longer exhibit significant changes over time, indicating the system had reached a steady state. The propulsion velocity is fixed at \( v_0 = 0.1 \), and the packing fraction is 0.3. We fix the translational and rotational noise strength $D_t= D_r=5\times 10^{-5}$. Simulations with \( N = 1000 \) to \( 16,000 \) particles confirm negligible finite-size effects, with \( N = 1000 \) used in subsequent results. For the reader's convenience, Table 1 provides a summary of the key parameters of our model and the computer simulations.

\begin{table*}[h!]
    \centering
    \caption{List of model parameters used in the simulations.}
    \begin{tabular}{llll}
        \hline \hline
        \textbf{Parameter}                 & \textbf{Notation}  & \textbf{Value}   & \textbf{Dimension} \\
        \hline
        Simulation time step               & $\Delta t$        & $10^{-2}$        & $1/(\mu k)$        \\
       Particle diameter                  & $d$                & 1                 & $d$                \\
        Translational diffusion coefficient & $D_T$              & $5 \times 10^{-5}$ & $d^2 \mu k$      \\
        Rotational diffusion coefficient    & $D_R$              & $5 \times 10^{-5}$ & $\mu k$        \\

        External potential parameter (strength) & $a$               & $1/30$           & $k_B \mathcal{T}$         \\
        External potential parameter (strength) & $b$               & $1/1.5$          & $k_B \mathcal{T}$         \\
        External potential parameter (asymmetry) & $c$              & Variable         & $k_B \mathcal{T}$         \\
        Packing fraction                   & $\phi$            & 0.3              & 1                  \\
        Propulsion velocity                & $v_0$              & 0.1              & $d \mu k$        \\
        Number of particles                & $N$                & 1000      & 1                  \\
        \hline \hline
    \end{tabular}
    \label{tab:parameters}
\end{table*}


\begin{figure}[h]%
\centering
\includegraphics[width=0.45\textwidth]{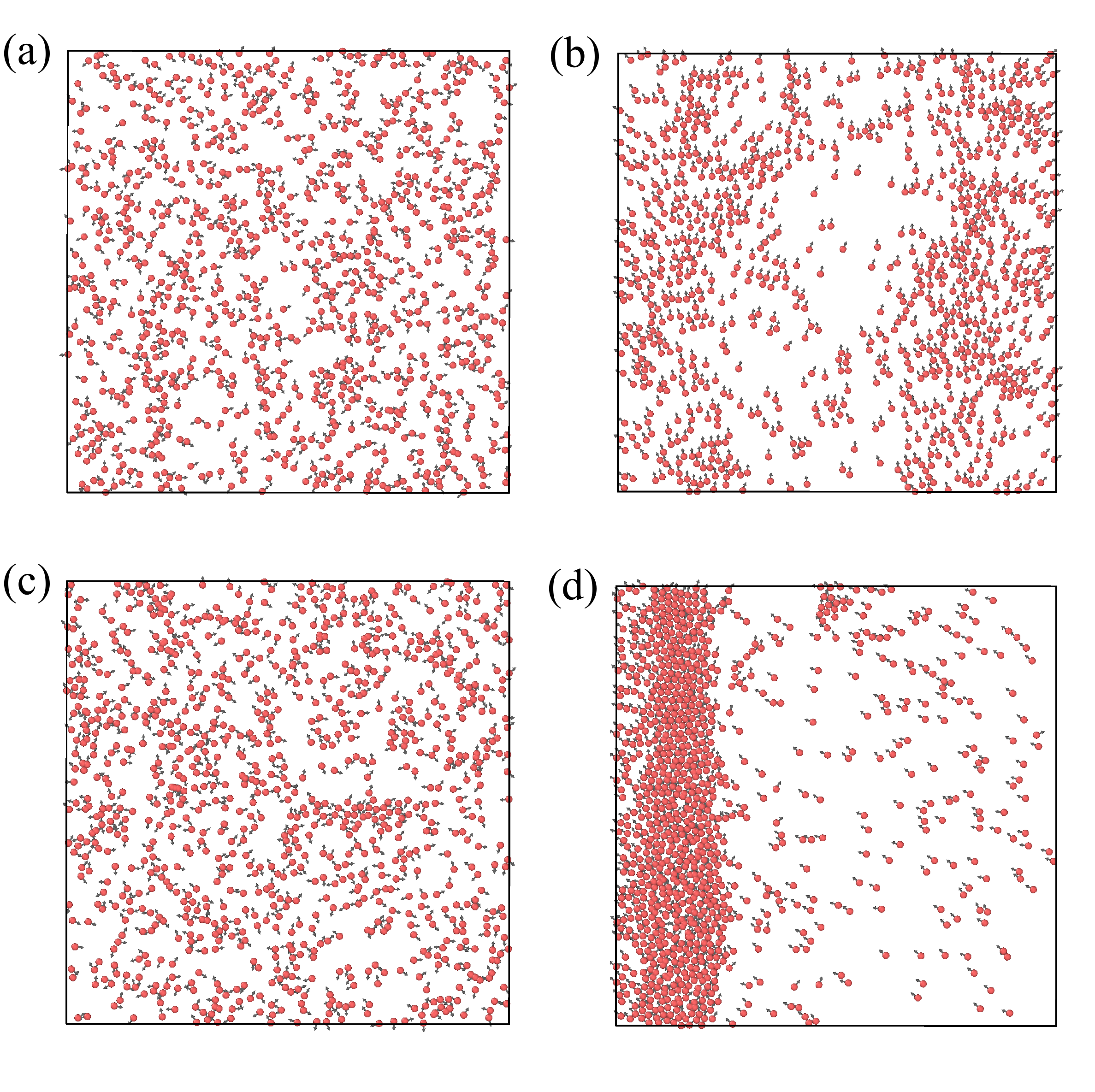}
\caption{Snapshots of the active system in the initial state and steady state. (a) initial state (disordered phase) with $c$ = 0.
(b) steady state (flocking) with $c$ = 0. (c) initial state (disordered phase) with $c$ = -0.5. (d) steady state (flocking and clustering) with $c$ = -0.5. The arrow represents the direction of the polarity. ($\alpha = 0.02$ for all figures).
}
\label{illustrate1}
\end{figure}

Fig.\,\ref{illustrate1} shows snapshots of the system at its initial and steady states. In a symmetric potential (\( c = 0 \)), particles transition from a disordered initial state [Fig.\,\ref{illustrate1}(a)] to a clear flocking steady state [Fig.\,\ref{illustrate1}(b)], where all particles move cohesively in the same direction, forming a symmetric distribution along the \( x \)-axis. In contrast, in an asymmetric potential (\( c = -0.5 \)), particles initially exhibit a homogeneous distribution [Fig.\,\ref{illustrate1}(c)] but evolve into a state characterized by flocking and clustering near the deeper side of the potential well [Fig.\,\ref{illustrate1}(d)]. This indicates that potential asymmetry drives spatially uneven particle distributions, ultimately leading to the formation of clustered states.

\begin{figure}[t]%
\centering
\includegraphics[width=0.5\textwidth]{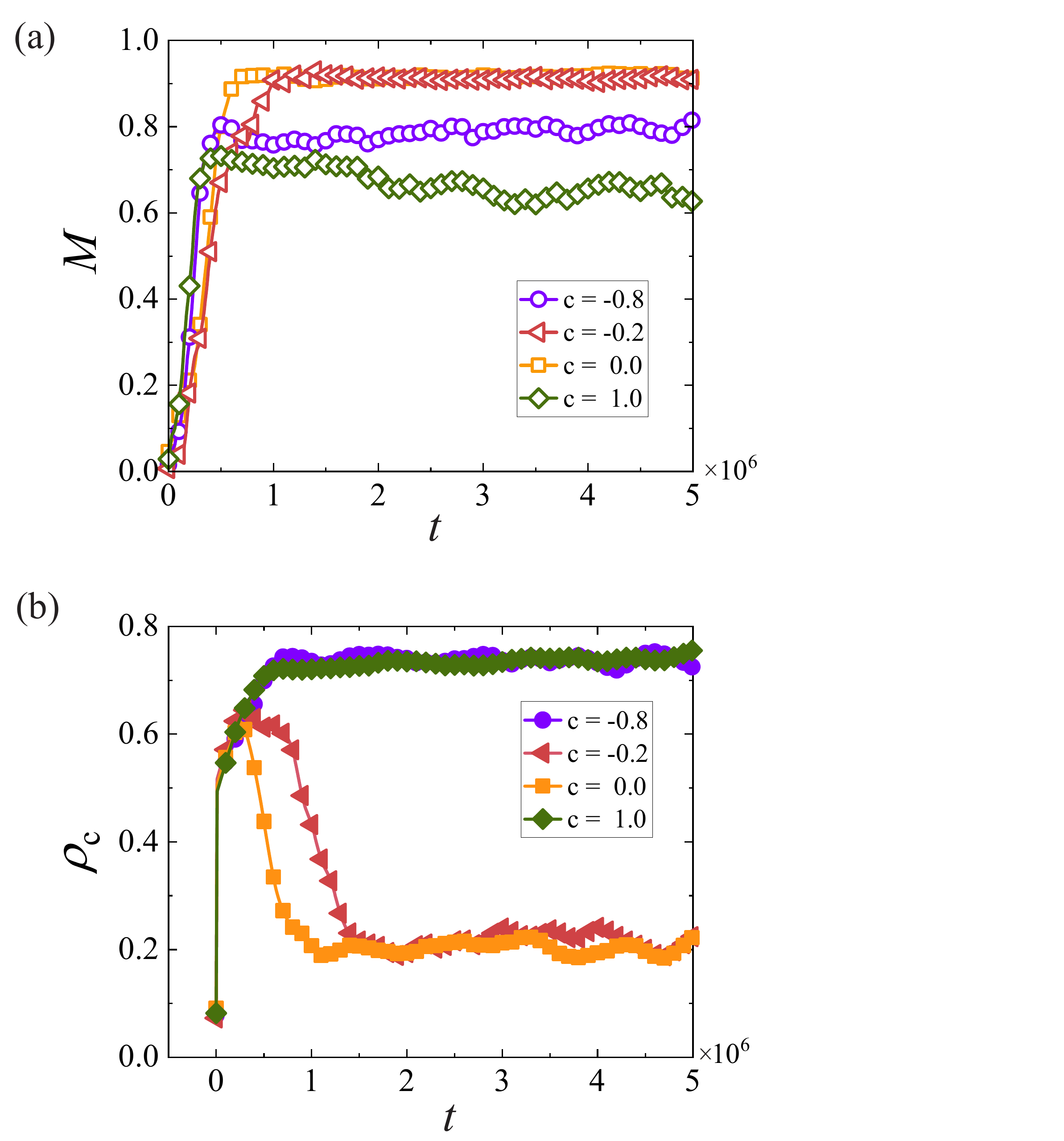}
\caption{(a) $M$ and (b) $\rho_c$ as a function of $t$ under different values of $c$. Simulation parameters: $\alpha$ = 0.02.}
\label{M-and-roc-vs-t}
\end{figure}
\section{phase behaviors}
We find that flocking and clustering appeared in the stability of system evolution, which did not appear in the previous studies. So based on the findings above, to quantify these effects and characterize the behaviors of phases in the system, we introduce order parameters. For flocking, we measure the net velocity, which is expressed as $M=\sqrt{\langle\cos \theta\rangle^2+\langle\sin \theta\rangle^2}$, where the average $\langle\rangle$ is taken over all $N$ particles. When $M$ approaches 0, it implies that the system is in a disordered state, and when $M$ approaches 1, it indicates that the system is in a highly ordered state. To quantitatively identify the clustering phase of the system that separates into dense clustered and dilute gas-like states, we measure the local area density of each particle~\cite{rodriguez2020phase}. The local area density of each particle is the ratio of the area of each particle to the area of the polygon via a Voronoi tessellation of each particle. By numerical simulation, in this work, we set the threshold of local area density as 0.6. We introduce the order parameter $\rho_c$ to describe clustering, which is the sum of the fraction of particles with its local area density larger than 0.6. Its math expression is shown as:
$\rho_c =\sum \rm{P}\left(\phi_l> 0.6\right)$.

We execute with 100 initial positions in the simulation to get average results. In order to explore the effect of potential asymmetry on phase behaviors, we calculate the dependence of $M$ and $\rho_c$ on $t$ under different values of $c$ in Fig.\,\ref{M-and-roc-vs-t}. Fig.\,\ref{M-and-roc-vs-t}(a) shows that with the increase of $\lvert c \rvert$, $M$ in the steady state would gradually decrease. Subsequently, we test the evolution of $\rho_c$ over time under the same condition above as shown in Fig.\,\ref{M-and-roc-vs-t}(b). When $\lvert c \rvert$ is relatively small, the system initially exhibits the aggregation of particles, gradually accumulates saturation to a larger scale cluster, and then gradually dissipates. However, when $\lvert c \rvert$ is large enough, the particles in the system gradually gather into a large cluster and no longer dissipate. By combining these two figures, it can be found that there is obvious flocking in the steady state without aggregation in the symmetric potential well ($c=0$). And when $\lvert c \rvert$ is large, the steady state of the system exhibits both clustering and flocking state. This is consistent with what is shown in Fig.\,\ref{illustrate1}. These results demonstrate a negative correlation between $M$ in the steady state and $\lvert c \rvert$ and a positive correlation between $\rho_c$ in the steady state and $\lvert c \rvert$.

\begin{figure}[t]%
\centering
\includegraphics[width=0.50\textwidth]{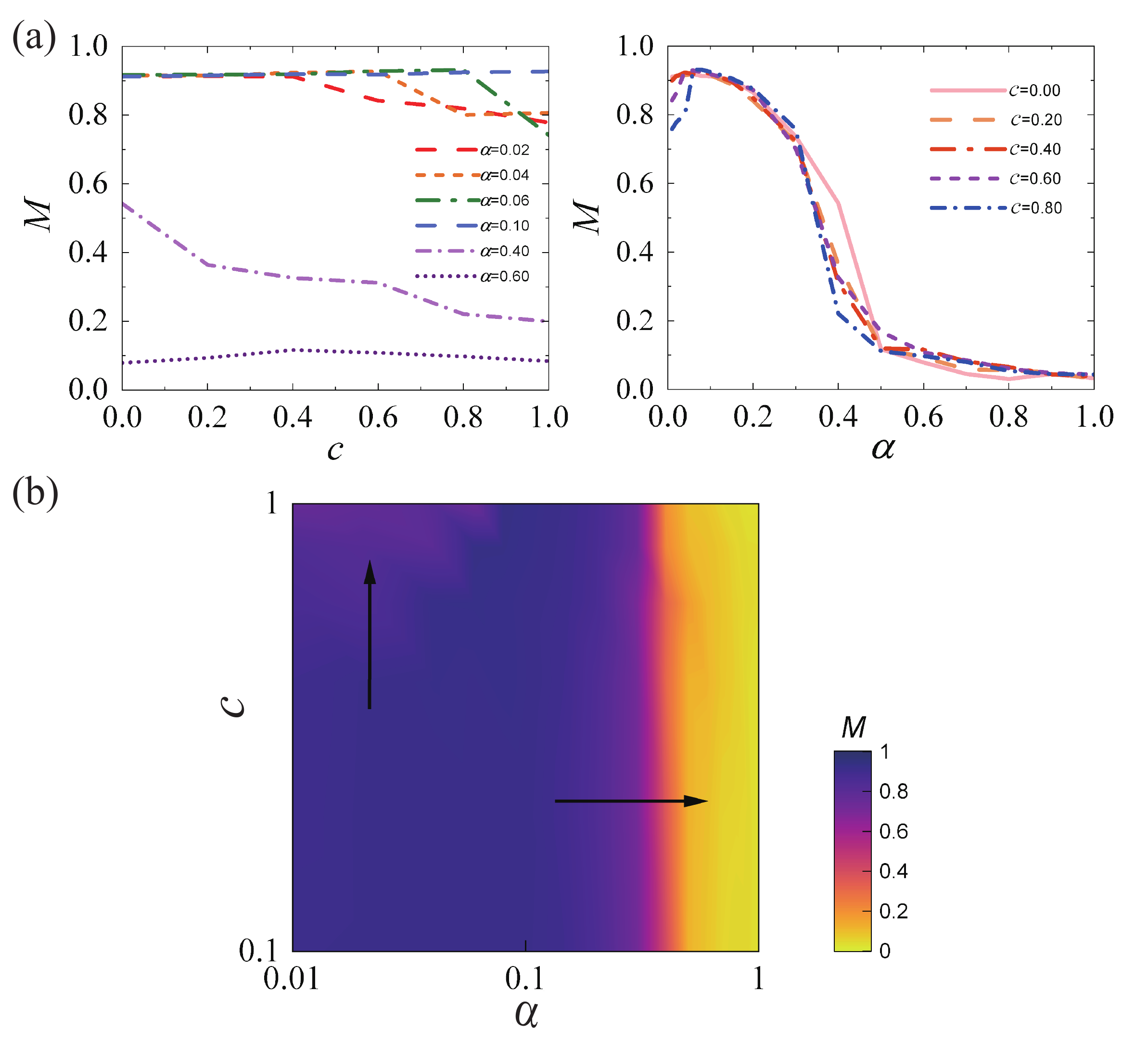}
\caption{(a) Flocking order parameter $M$ versus $c$ and $\alpha$ as denoted in the key. (b)Simulations reveal dependence on both active reorientation $\alpha$ and asymmetry parameter $c$. All data was averaged over time in a steady state.}
\label{contourmap}
\end{figure}

Next, we explore the steady state of the system, focusing on flocking. Fig.\,\ref{contourmap}(a) illustrates two relationships: how $M$ changes with $c$ at fixed values of $\alpha$ and how $M$ changes with $\alpha$ at fixed values of $c$. The results reveal that $M$ is more sensitive to changes in $\alpha$ than to changes in $c$. Specifically, for different values of $c$, $M$ decreases sharply as $\alpha$ increases and approaches 0.4. When $\alpha$ is small, $M$ gradually decreases as $c$ increases. However, for larger $\alpha$, the influence of $c$ on $M$ becomes negligible, indicating that activity dominates over asymmetry in determining the system's behavior. These findings suggest a competitive relationship between activity and asymmetry in this system. To further explore the competitive relationship between the two, we test the phase diagram of flocking about $c$ and $\alpha$, where the color represents the order parameter of flocking $M$ in Fig.\,\ref{contourmap}(b). It is evident that when $\alpha$ is small, the larger the asymmetry $c$ of the potential, the smaller $M$. When $\alpha$ is large, activity becomes the dominant factor. $\alpha = 0.4$ can be considered a critical $\alpha$, serving as the boundary between ordered and disordered states of the system. It is also clear that changes in activity have a more significant impact on $M$ compared to the degree of asymmetry.

\section{Dynamics}
\begin{figure}[t]%
\centering
\includegraphics[width=0.45\textwidth]{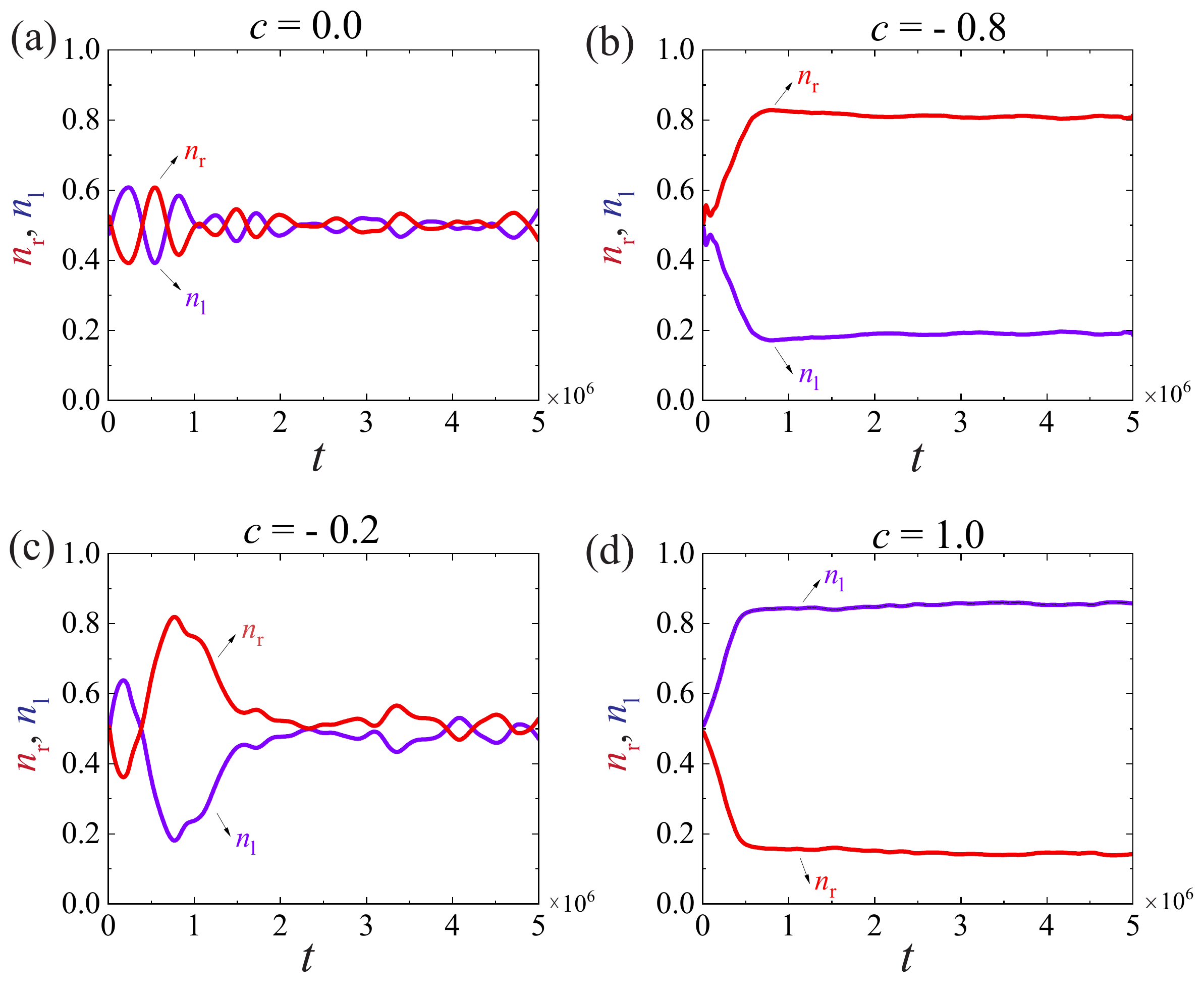}
\caption{$n_{l}$ and $n_{r}$ as a function of time for various values of $c$ in this active system with active reorientation ($\alpha=0.02$).}
\label{number-of-particles-evolve}
\end{figure}

\begin{figure}[t]%
\centering
\includegraphics[width=0.45\textwidth]{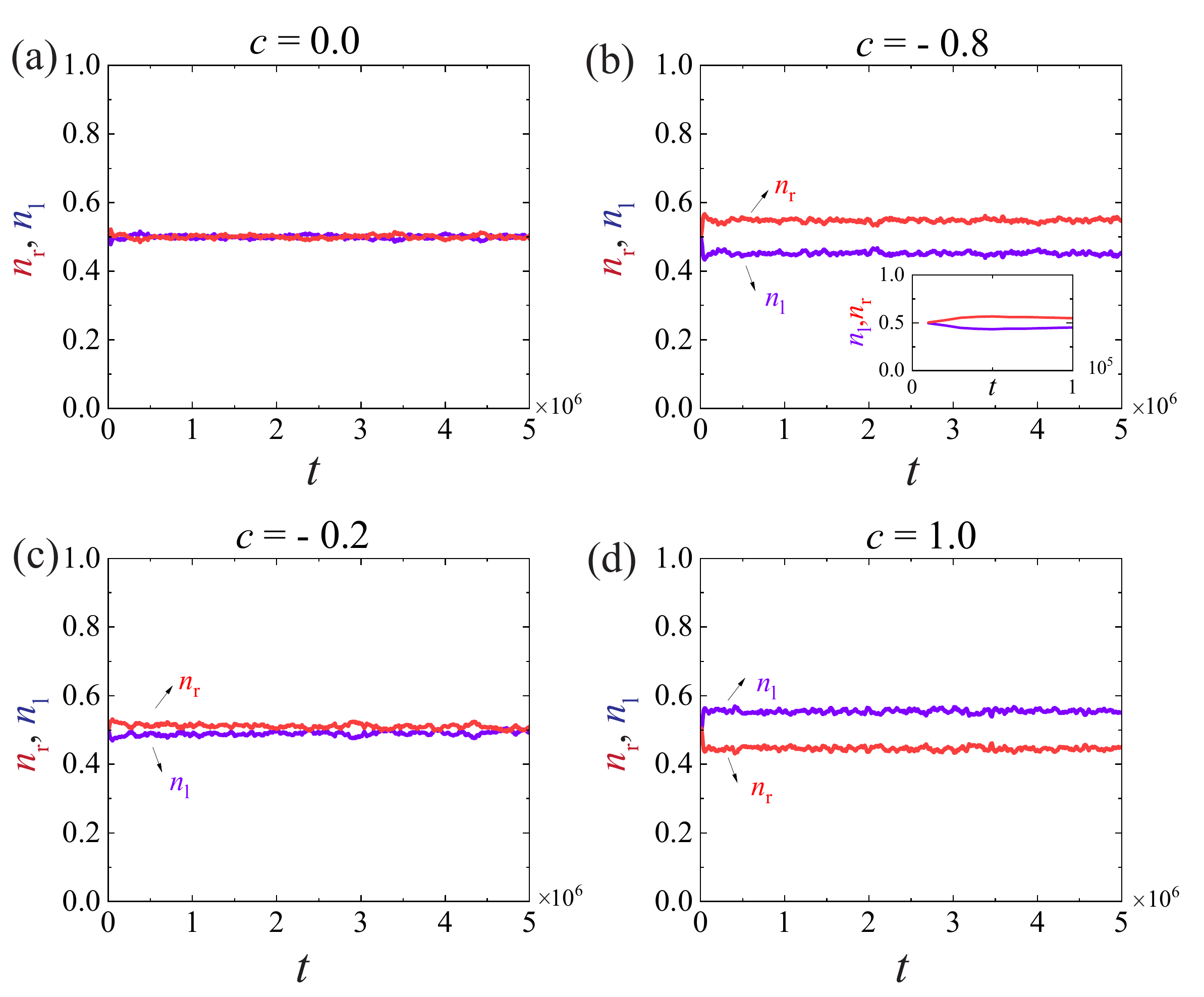}
\caption{$n_{l}$ and $n_{r}$ as a function of time for various values of $c$ in this active system without active reorientation ($\alpha=0$).}
\label{non-AR}
\end{figure}

The dynamics of the system are studied by analyzing the fraction of particles in the left and right wells in the simulation domain. Here we use $n_{l}$ and $n_{r}$ to represent the number of particles in left and right well, respectively. Fig.\,\ref{number-of-particles-evolve} presents the dependence of $n_{l}$ and $n_{r}$ on time series in the symmetric and asymmetric potentials. In Fig.\,\ref{number-of-particles-evolve}(a), when $c$ is 0, we find that the number of particles distributed in the left and right x-axis at the first stage of evolution shows a state of periodic exchange between the left and right positions over time, and then particles gradually distribute on both sides equally tending to be stable. The number of particles in the potential on the left and right of the x-axis is the same dynamically in the symmetric potential wells, which suggests the system does not exhibit directed motion in the x-axis in a symmetric potential. This is different from the situation where $\lvert c \rvert$ is larger. As we can see from Fig.\,\ref{number-of-particles-evolve}(b), when the parameter of asymmetry of the potential is -0.8, we find that particles in this potential gradually evolve from the uniform state in the position space to a phase separation state. It is shown that the particles are confined by the potential field in the lower region of the potential well. On the other side, there are essentially few particles. At the same time, we observe that when $c$ is 0.2 as shown in Fig.\,\ref{number-of-particles-evolve}(c), there seems to be a situation where the left and right sides exchange with each other just like the results with $c$ is 0, but the exchange frequency is not as obvious as Fig.\,\ref{number-of-particles-evolve}(a). Compared with Fig.\,\ref{number-of-particles-evolve}(b), Fig.\,\ref{number-of-particles-evolve}(d) shows that the degree of particle exchange at the left and right potential gradually decreases with the increase of $\lvert c \rvert$. These results indicate that the parameter of asymmetry in the potential field plays a crucial role in determining the capture of particles.
\begin{figure}[t]%
\centering
\includegraphics[width=0.4\textwidth]{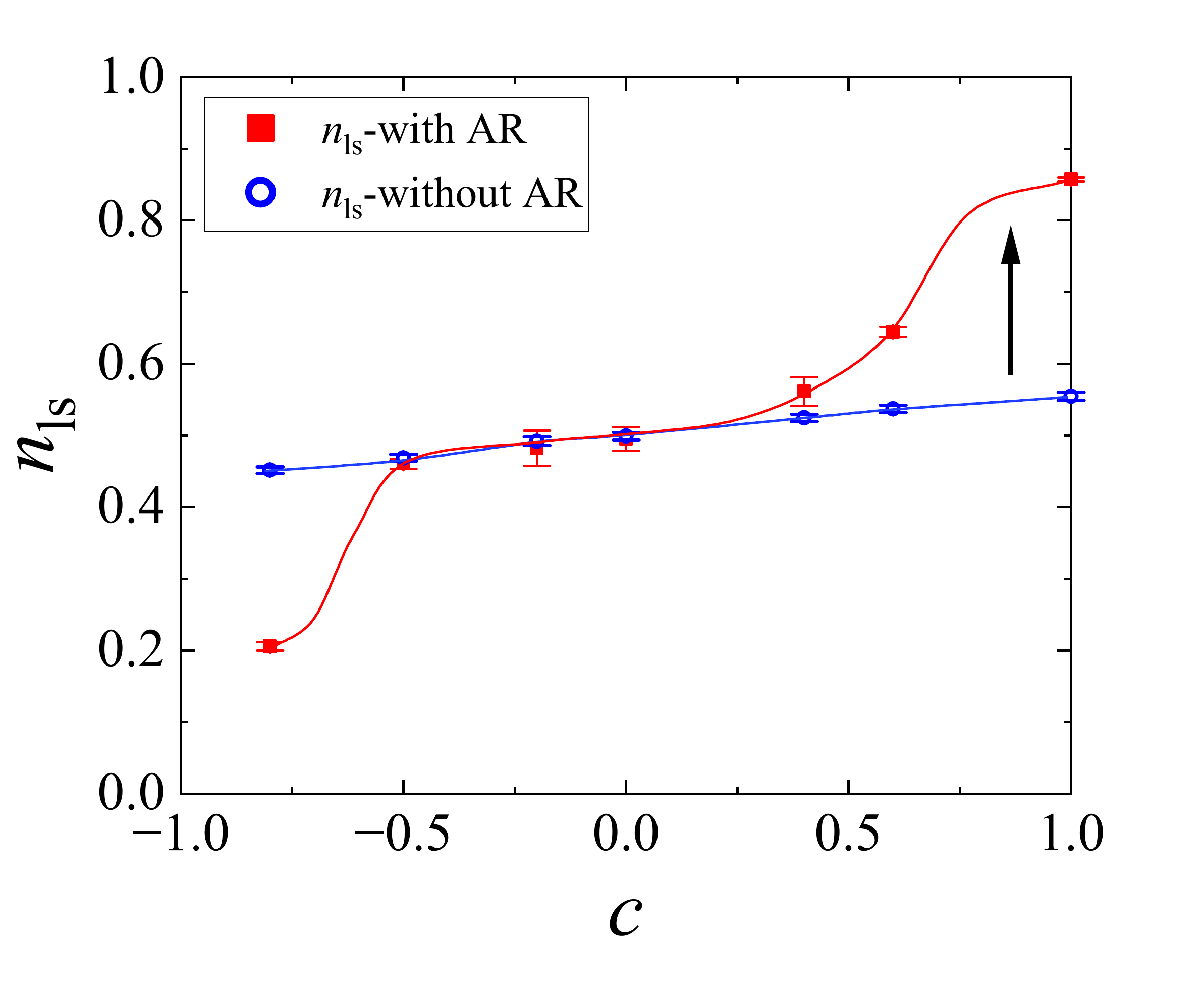}
\caption{A plot on the parameter of asymmetry $c$ dependence of $n_{l}$ evaluated at steady state.}
\label{illustrate3}
\end{figure}
Next, as a comparison, we explore the dynamics of an active system without active reorientation in the double-well potential which is shown in Fig.\,\ref{non-AR}. First of all, we notice that the distribution of particles in the left and right sides of the potential is uniform with $c=0$ in Fig.\,\ref{non-AR}(a) just like the results above in Fig.\,\ref{number-of-particles-evolve}(a).
As the value of $\lvert c \rvert$ increases, the particles also exhibit a directed flow towards the deeper side of the potential field. Notably, the gap between $n_{l}$ and $n_{r}$ is smaller than the gap in the active system with active reorientation. This also indicates that active reorientation enhances the directed motion of the system in the asymmetric potential. In other words, active reorientation endows the system with a stronger ability for directed leadership compared to a system lacking active reorientation.

Then we average $n_{l}$ in the Fig.\,\ref{number-of-particles-evolve} and Fig.\,\ref{non-AR} of time after $3\times 10^{6}$ and obtain $n_{ls}$ as the number of particles in the left potential in the steady state and standard deviation of $n_{ls}$. Fig.\,\ref{illustrate3} presents $n_{ls}$ as a function of different values of $c$. We observe that $n_{ls}$ exhibits an anti-symmetric form concerning $c=0$. Diamond points shown in Fig.\,\ref{illustrate3} demonstrate that when the value of $\lvert c \rvert$ is less than 0.5, the steady state of the system demonstrates relatively uniform on both sides. When $\lvert c \rvert$ is greater than 0.5, the steady-state exhibits an obviously asymmetric state with particles concentrating in the deeper side region. As for particles without active reorientation(circle points), the behavior of $n_{1s}$ with $c$ is consistent with that of the active reorientation case, but the degree of change is much smaller. It demonstrates that the amplification of active reorientation enhances the effect of asymmetry on the distribution of particles along the x-axis.

\section{Different diffusion regimes}
\begin{figure}[t]%
\centering
\includegraphics[width=0.50\textwidth]{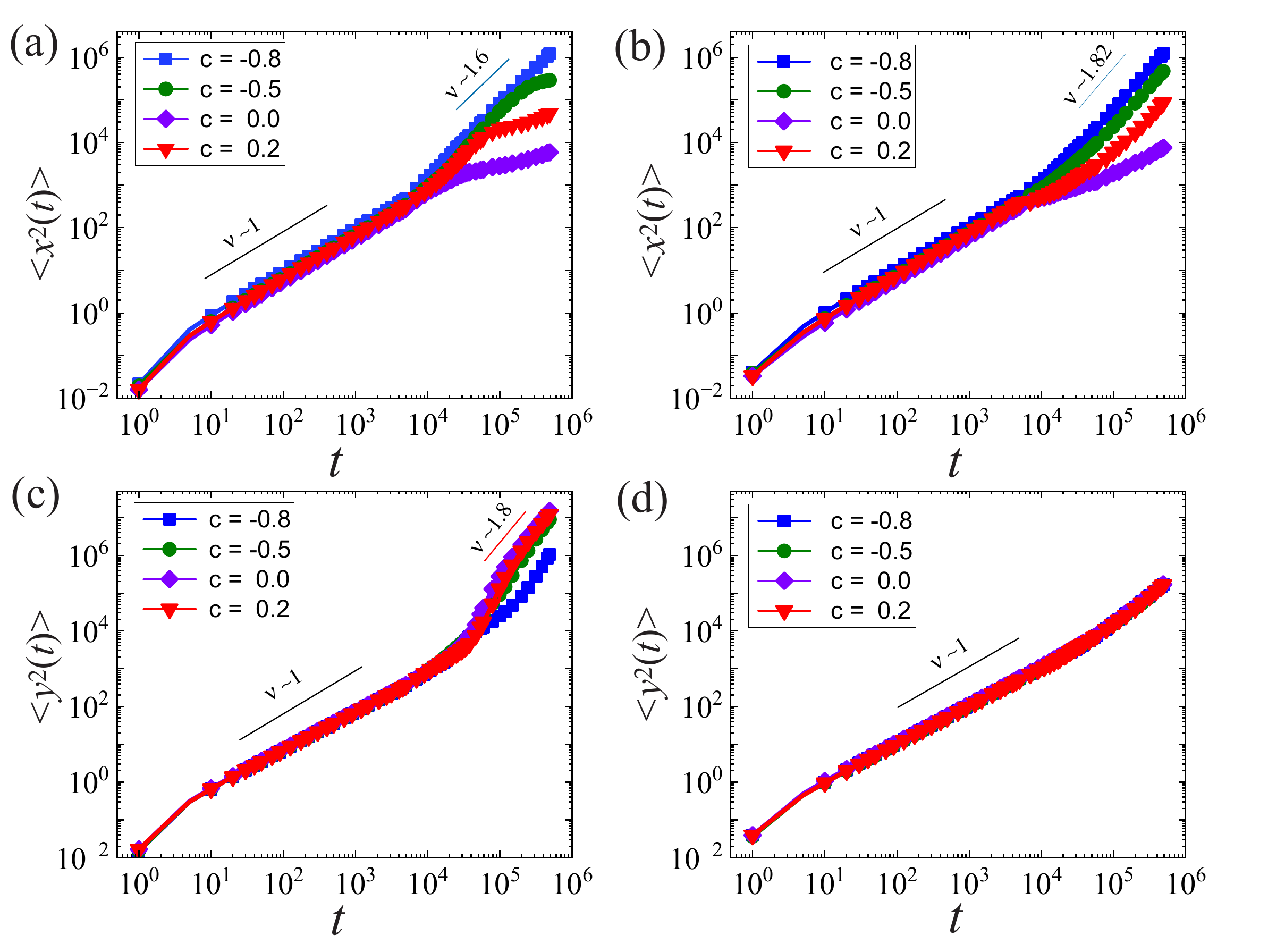}
\caption{The MSDs along the x and y directions versus time for various values of $c=-0.8,-0.5,0.0,0.2$. (a) MSDs in the x-direction of the system with and (b) without active reorientation. (c) MSDs in the y-direction of the system with and (d) without active reorientation. The solid lines represent the exponential fittings.
}
\label{MSD-diff-C}
\end{figure}

The investigation of diffusive behavior has a long and rich history. In 1920 Fuerth first discussed Brownian motion with a consideration of persistence in motion direction~\cite{furth1920brownsche}, extending the classical works of Smoluchowski and Einstein. This pioneering work established the foundation for understanding deviations from purely random motion due to persistence effects, which are crucial in describing modern active particle systems. Specifically, Fuerth introduced the idea of directional persistence influencing mean-squared displacement, which has since been expanded to encompass active diffusion driven by self-propulsion. In the active system of self-propelled particles moving in a double well potential, the dynamics can be influenced by both the underlying potential and the self-propulsion as well as the active rotation of the particles. So next we focus on the long time dynamic behaviors when the asymmetry parameter of potential increases by calculating the mean-squared displacement (MSD), which is
\begin{equation}
\langle r^{2}(t)\rangle=\frac{1}{N}\sum_{i=1}^{N}<[r_{i}(t)-r_{i}(0)]^{2}>
\end{equation}
where $r_i$ is the position of particle $i$, and $N$ is the total number of particles in the studied system. As we know, the relationship between the MSDs and time $t$ can be given by expressed as:

\begin{equation}
\langle r^2(t) \rangle \sim t^\nu
\end{equation}
where $\nu$ is the scaling exponent. So we test MSD and get $\nu$ in the x and y directions under different values of cc both in the system with AR (Fig.\,\ref{MSD-diff-C}(a) and (c))and without AR (Fig.\,\ref{MSD-diff-C}(b) and (d)). Fig.\,\ref{MSD-diff-C} shows that there is a clear difference in scaling behavior between particles with active reorientation and without active reorientation of MSD. First, we focus on the MSD of the system with AR in Fig.\,\ref{MSD-diff-C}(a). We found that similar to the passive system, a brief period of diffusive motion is observed initially, followed by a transition to a subdiffusive regime due to the caging effect. While the subdiffusive phase consistently emerges, the subsequent superdiffusive behavior—known in the context of free active particles—is not universally observed. Its occurrence depends on the interplay between structural relaxation and the simultaneous decorrelation of rotational degrees of freedom~\cite{reichert2021transport,sharma2016communication}.

Next comparing MSDs of AR and no AR systems, in the short time scale, both AR and no AR systems exhibit Brownian motion under different degrees of asymmetry. In passive systems, this subdiffusive behavior is often attributed to particle trapping within potential wells, where the thermal fluctuations play a dominant role in overcoming energy barriers, resulting in slower-than-diffusive motion. However, in the long time scale, it is found that in the x direction (Fig.\,\ref{MSD-diff-C}(a) and (b)), the caging effect of the system with AR is significantly stronger (as indicated by a lower subdiffusion exponent) than that of the system without AR under the same asymmetry. This behavior is consistent with some previous findings. In the previous work~\cite{Reichert2021EPJE44}, researchers claim that the exponents $\nu$ of the free particle consistently reveal the occurrence of caging in the form of pronounced minima and the enhancement of superdiffusion with increasing persistence length. Further, they confirm that superdiffusive behavior can be suppressed in the entire time window which is the case under certain packing fractions and translational diffusive coefficients. And as we know Rouse theory~\cite{Doi1986Oxford} predicts significant subdiffusive behavior of the monomer mean square displacement (MSD) due to chain connectivity in the limit of long chains, with MSD $\sim t^{0.5}$. The work by Li et al~\cite{zhangbk2024softmatter} found activity in the unentangled semidilute polymer raises the ambient temperature, which significantly decreases diffusion by over an order of magnitude. But in this paper, activity is manifested as active reorientation capability, rather than activity enhanced by increasing temperature. In addition, Kumar et al.~\cite{kumar2023dynamics} simulated tracer dynamics in a diamond-lattice polymer network, demonstrating that active self-propulsion induces intermediate-time subdiffusive motion for larger sticky tracers and pronounced superdiffusive behavior with non-Gaussian characteristics for smaller, highly active tracers, with transport dynamics strongly modulated by the network stiffness. The work of Shin et al.~\cite{shin2017elasticity} examined polymer dynamics in active media, demonstrating superdiffusive motion at intermediate times and greatly enhanced diffusivity at longer times, with non-monotonic behavior influenced by polymer stiffness and chain length. They further explored how directed motion arises from the interplay of chain elasticity and active forces. Goswami et al.~\cite{goswami2024anomalous} studied the dynamics of self-propelled particles in deformable gel-like structures, observing transitions from subdiffusion to superdiffusion driven by trapping and escape mechanisms. Their work revealed nonergodic behavior and complex waiting time distributions dependent on activity levels, complementing our analysis of diffusion regimes in double-well potentials.

\begin{figure}[h]%
\centering
\includegraphics[width=0.50\textwidth]{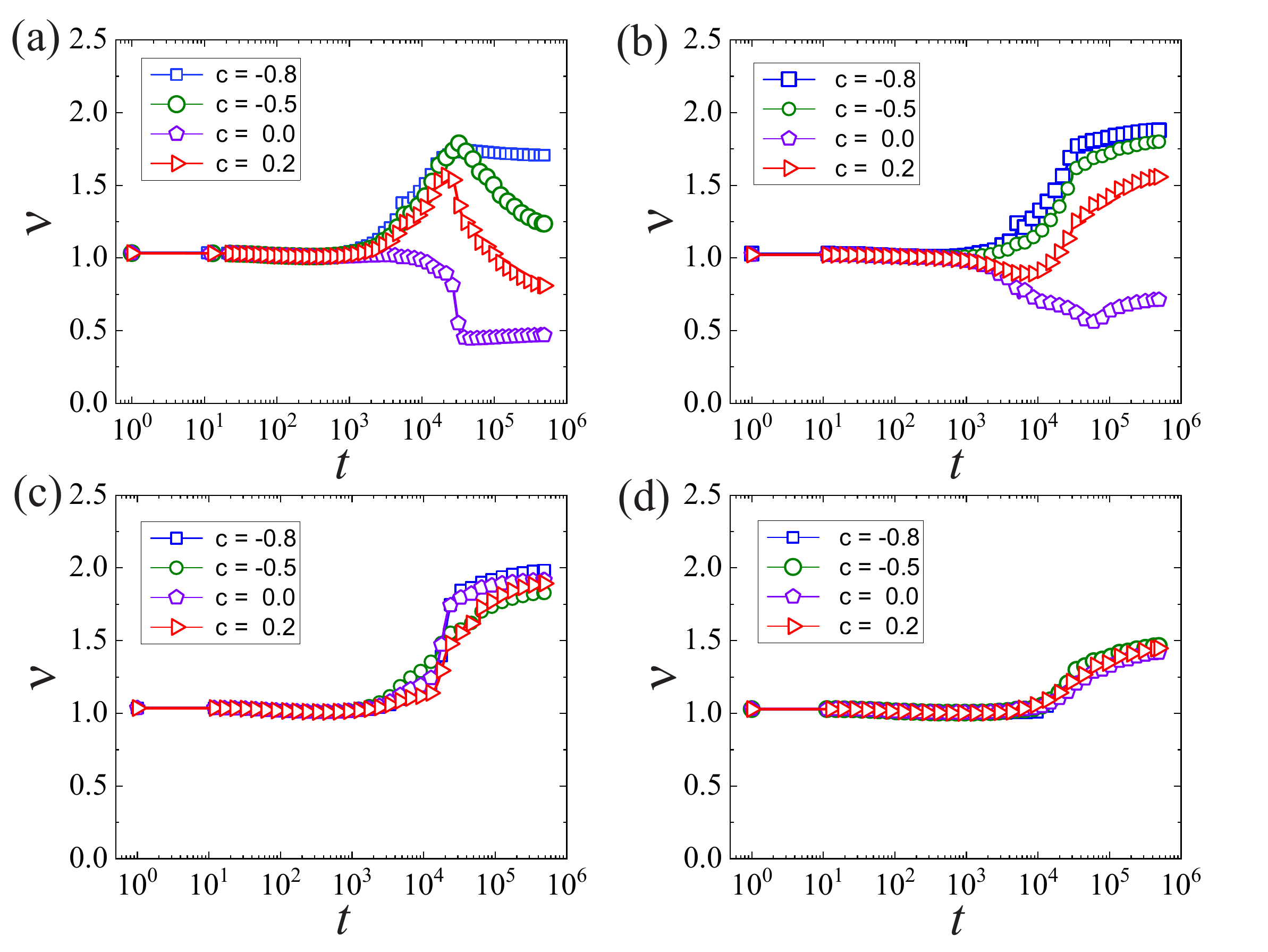}
\caption{The local exponents as a function along the x and y directions versus time for various values of $c=-0.8,-0.5,0.0,0.2$. (a) Local exponent in the x-direction of the system with and (b) without active reorientation. (c) Local exponent in the y-direction of the system with and (d) without active reorientation.
}
\label{figlocal-exponent}
\end{figure}
In the y direction (Fig.\,\ref{MSD-diff-C}(c) and (d)), the situation is reversed. In the long time scale, the active system exhibits superdiffusive behavior under the same asymmetry, with higher diffusion intensity compared to the passive system. This is a particularly interesting finding.
The reason is that mobility-induced phase separation which results from active particles with repulsive force causes the coupling between self-propelling and potential. In contrast, passive systems do not display this type of collective dynamic transition, as their motion remains dominated by thermal fluctuations rather than directed propulsion. In the meantime, according to previous results~\cite{PhysRevLett.108.235702, PhysRevLett.110.055701}, steady states of active systems with self-propulsion ability do not exhibit flocking behavior but mobility-induced phase separation. However, our system due to active reorientation displays collective flocking motion in the steady state when values of $\alpha$ and $D_r$ are relatively small~\cite{chen2023initial}. Furthermore, during this steady state, the system no longer shows clustering~\cite{chen2023initial}. Now, when an asymmetric potential trap is introduced, it significantly confines the active particles to one side. These active particles must interact closely while being confined together, continuously reorienting themselves. This, in turn, enhances the probability of particle overlap and reorienting, leading to a relatively quicker attainment of the flocking state or in other words, quickly reaching a stable state(see Fig.\,\ref{M-and-roc-vs-t}(a)). Finally under the same parameter $c$, there seems to be no significant difference in the x-direction between systems with active reorientation and those without active reorientation. However, in the y-direction, the system with active reorientation exhibits much stronger superdiffusive behavior compared to the system without active reorientation. The difference in MSD in the y direction also suggests that the potential in this direction is more complex or restrictive, making active reorientation a key factor in enhancing particle mobility.

To reveal the dynamic characteristics and transitions across different time scales, we introduce the local scaling exponent. The local scaling exponent quantitatively describes the scaling behavior of the mean squared displacement (MSD) over time, which can be determined using the following formula~\cite{metzler2014anomalous}:

\begin{equation}
\nu(t)=\partial \log \left\langle{r^2(t)}\right\rangle / \partial \log t
\end{equation}
The slope $\nu(t)$ in logarithmic space allows us to extract the scaling exponent at different time points, enabling us to identify whether the system exhibits superdiffusion ($\nu \textgreater 1$), normal diffusion ($\nu=1$), or subdiffusion ($\nu \textless 1$). Based on this approach, we plot the variation of the local scaling exponent with time (as shown in Fig.\,\ref{figlocal-exponent}). Fig.\,\ref{figlocal-exponent}(a) and (c) illustrate the local scaling behavior in the x and y directions under the influence of AR, while Fig.\,\ref{figlocal-exponent}(b) and (d) correspond to the cases without AR. In the x direction, when the time is less than $10^3$, the system exhibits normal diffusion. However, as time increases, the system shows superdiffusion when the asymmetry parameter $c \neq 0$, whereas it exhibits subdiffusion when $c = 0$. Additionally, the local scaling exponent increases with the absolute value of c. In the y direction, the critical time for the transition from normal diffusion to superdiffusion is approximately $10^4$. Comparing Fig.\,\ref{figlocal-exponent}(b) and (d) reveals that AR significantly enhances the diffusion capability of the system.

\section{Summary and conclusions}
In this study, phase behaviors and dynamics of active systems with AR in the double-well potential are investgated. We obtain the phase diagram as a function of the asymmetry parameter of the potential and active reorientation. By comparing systems with and without AR, we find that the system with AR amplifies the effect of the asymmetry of potential, enhancing the directed motion for active particles. Finally, by calculating the mean squared displacement and local scaling exponent over time, we highlight the distinct time scales over which different diffusion regimes occur, revealing clear crossovers among subdiffusion, normal diffusion, and superdiffusion. These findings demonstrate the effect of potential on the dynamic behavior of various physical nonequilibrium systems with active reorientation.

\section*{Acknowledgements}
We wish to acknowledge Z. C. Tu offering suggestions. We also acknowledge computational support from the Beijing Computational Science Research Center and Changchun Normal University. The research was supported by the Natural Science Foundation of Changchun Normal University.

{}

\end{document}